\documentclass[runningheads]{llncs}
\usepackage[T1]{fontenc}
\usepackage{graphicx}
\usepackage{enumitem}
\usepackage{amsmath}
\usepackage[ruled,linesnumbered]{algorithm2e}
\usepackage{siunitx}
\usepackage{booktabs}
\usepackage{cite}
\usepackage{subcaption}
\usepackage{hyperref}
\usepackage{multirow}
\usepackage{bbding}

\usepackage{color}

\begin{document}
\title{SET: Stream-Event-Triggered Scheduling for Efficient CUDA Graph Pipelines}
\titlerunning{SET}
% If the paper title is too long for the running head, you can set an abbreviated paper title here

\author{Zhengxiong Li\inst{1}\Envelope \and
Tsung-Wei Huang\inst{1} \and
Umit Ogras\inst{1}}
\authorrunning{Z. Li, et al.}
\institute{University of Wisconsin-Madison, Madison WI, 53706, USA \\
\email{\{zhengxiong.li, tsung-wei.huang, uogras\}@wisc.edu}\\
}
\maketitle              % typeset the header of the contribution
\vspace{-5mm}
\begin{abstract}
Achieving peak GPU performance remains a significant challenge as the system throughput is constrained by host-device synchronization delays and kernel scheduling overheads, even with aggressive kernel optimizations and batch processing. Furthermore, existing approaches often underutilize hardware resources such as compute cores and copy engines due to scheduling overheads.
To address these problems, 
we propose a CUDA runtime framework for task-parallel pipelines to minimize the synchronization overheads and the gap between kernel executions. 
The proposed solution combines two innovations: (1) a multi-stream task-parallel pipeline programming model that leverages event-chaining and work-stealing mechanisms to fully utilize available hardware resources; (2) a graph-based execution flow with per-stream buffers to ensure memory safety for multiple in-flight jobs running concurrently. 
Extensive evaluations on representative real-world workloads show 1.15--1.44$\times$ speedup and reduce scheduling overheads by 18--54\% compared to state-of-the-art CUDA graph baselines.

\keywords{CUDA Graphs \and GPU Runtime \and Dynamic Task Scheduling.}
\end{abstract}

\section{Introduction}
\label{sec:intro}

Modern GPU applications, ranging from AI and machine learning to scientific and image processing workloads, execute as pipelines of interdependent kernels connected by data transfers and synchronization operations. To reduce the repeated cost of launching these kernels, NVIDIA introduced the CUDA Graph execution model, which captures a directed acyclic graph (DAG) of kernel launches and memory operations that can be efficiently instantiated and replayed~\cite{NvidiaCudaGuide}. CUDA Graph reduces the CPU-side launch overhead and improves throughput compared to traditional stream-based execution. It enables developers to achieve substantial performance improvements without optimizing memory access patterns and kernels.

\begin{figure}[htbp]
    \centering
    \includegraphics[width=0.8\linewidth]{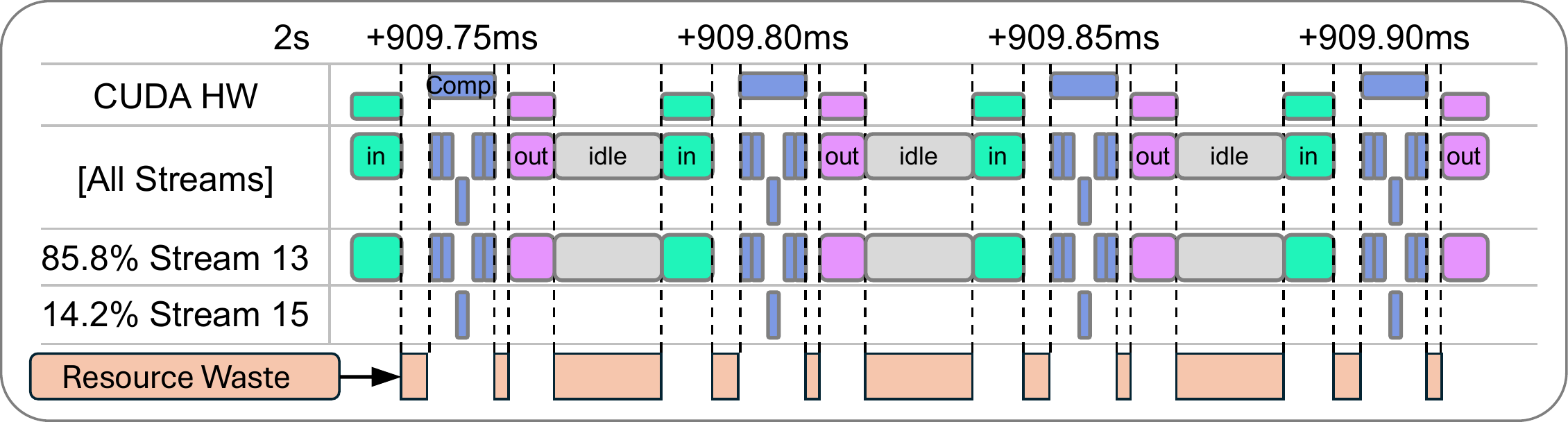}
    \vspace{-2mm}
    \caption{Gaps between operations in Nsight profiler}
    \label{fig:profiler}
    \vspace{-8mm}
\end{figure}

Many real workloads fail to achieve full GPU utilization even with CUDA graphs~\cite{Zheng2023}. Idle intervals—called kernel gaps—often appear on the GPU timeline between successive kernel executions.
These gaps arise from various reasons, such as synchronization mechanisms and CPU-side operations, including argument updates, device buffer management, stream selection, and \texttt{memcpy} enqueuing. 
Nsight Systems Profiler's event timeline manifests these issues as recurring white spaces between the last instruction of one kernel and the first of the next, as shown in Fig.~\ref{fig:profiler}. Minimizing these kernel gaps is crucial to realizing the full performance potential of CUDA graph pipelines.

\vspace{-5mm}

State-of-the-art approaches often submit jobs in batches~\cite{Ekelund2025}. 
They launch CUDA graphs that represent a batch of multiple parallel processing chains, while the host synchronizes after each batch completes. 
While this method effectively parallelizes the GPU's computational tasks, it fails to address the underlying issue: CPU-GPU dependency. 
Specifically, this strategy assigns work to a fixed set of streams and relies heavily on host-side polling or blocking to safely reuse stream buffer space, avoiding overwriting outputs from prior kernels. As a result, copy engines are underutilized, and compute cores may remain idle when any stream has no work to do. 
Although CUDA graphs instantiation, limited batching, or stream callbacks attempt to reduce the average overhead, they fail to react to fine-grained readiness events or keep enough independent jobs in flight to absorb jitter. Hence, these approaches often fail to hide the CPU overheads or fully utilize hardware resources.

This paper proposes SET, a stream-event-triggered runtime scheduling framework for CUDA graphs that maximizes system throughput and minimizes idle gaps. SET co-schedules host and device tasks by representing each job as a reusable CUDA graph executable and coordinating multiple workers through event chaining. Rather than relying on batch-level synchronization or global polling, SET maintains multiple in-flight job queues and dynamically assigns work to streams as they become available. When a stream completes execution, event callbacks trigger resource release and enable the scheduler to promptly dispatch the next ready job. This event-driven design reduces host-side scheduling latency, mitigates inter-batch gaps, and improves hardware utilization while preserving per-job ordering and memory safety. As a result, SET sustains high throughput across workloads with diverse characteristics, including short-lived kernels and memory-bound tasks.

Our contributions in this paper are outlined as follows:
\vspace{-2mm}
\begin{itemize}[leftmargin=*]
    \item \textbf{Bottleneck and gap characterization}. We analytically decompose scheduling overheads in CUDA graph pipelines into intra-batch and inter-batch components and explain why kernel gaps persist despite graph instantiation and multi-stream execution.

    \item \textbf{Event-chained host–device co-scheduling}. SET achieves adaptive task assignment with O(1) synchronization overhead while preserving per-job ordering and memory safety by using bounded in-flight CUDA graph executables and callback-driven coordination to minimize inter-kernel gaps.

    \item \textbf{Comprehensive evaluation across diverse workloads}. We evaluate SET on six representative compute- and memory-bound workloads on NVIDIA RTX 3090 and RTX 5090, and demonstrate 1.15--1.44$\times$ speedup and 18–54\% lower scheduling overheads than state-of-the-art host-side scheduling baselines.

\end{itemize}

In the rest, Section~\ref{related_work} discusses the prior work, Section~\ref{sec:background} provides background and motivation, Section~\ref{sec:methodology} presents the proposed SET framework. Finally, Section~\ref{sec:evaluation} presents the experimental evaluation and Section~\ref{sec:conclusion} concludes the paper.

\vspace{-5mm}
\section{Related Work} \label{related_work}
\vspace{-2mm}

Prior work has explored three directions to minimize CUDA programming overheads: (1) \textit{host-side scheduling} targets how the jobs are launched; 
(2) \textit{domain-specific language (DSL) compilers and runtime engines} move scheduling/fusion decisions upstream; 
(3) \textit{device-side dispatchers} move kernel launching decision to the device to minimize the host-device synchronization overheads.

\noindent\textbf{Host-side scheduling} enables performance optimization without modifying the kernels and source code. \textit{The synchronous model}~\cite{NvidiaCudaGuide} launches kernels one by one on a single stream without any explicit scheduling optimization mechanism. \textit{The CUDA Graph model}~\cite{NvidiaCudaGuide} reduces CPU overheads by pre-recording directed acyclic graphs (DAGs). However, it still pays per-replay and per-iteration parameter-update overheads for \texttt{memcpy} and kernel nodes, with limited adaptivity. \textit{The static batching model}~\cite{Ekelund2025} aggregates multiple jobs into a large CUDA graph to amortize launch overheads across multiple tasks, improving throughput at the cost of increased latency and memory pressure. 
\textit{The queue model}~\cite{Guevara2009EnablingTP} dynamically balances workloads across workers. It smooths kernel gaps but incurs $O(b)$ host-side queries and can perform badly when kernels have the same order of time as the overheads of managing shared queues.
Since SET is most closely related to these models, Section~\ref{sec:evaluation} presents detailed comparisons to them under six workloads. Unlike these models, SET provides adaptive per-worker queues, bounded in-flight graphs which allow on-the-fly argument updates, and O(1) shared resources overheads via callback events to sustain overlap. 

\noindent\textbf{Device-side dispatching} aims to minimize the host interaction using persistent kernels or threads on the device. 
\emph{Whippletree}~\cite{Steinberger2014} and \emph{Softshell}~\cite{Steinberger2012} keep \emph{persistent kernels} on the GPU pulling tasks from on-device queues. By scheduling work directly on the streaming multiprocessors (SMs), they reduce host launch overhead and improve fine-grained load balance. 
However, they cause fairness, backpressure, and termination detection challenges, which complicate interoperability with vendor libraries.
\emph{Kernelet}~\cite{Zhong2014} time-slices and co-schedules small kernels to increase SM utilization for irregular workloads. 

\noindent\textbf{DSL compilers and task runtimes} offer an orthogonal direction to reduce the number of kernels that need to be launched. They also improve memory utilization since merged kernels reuse intermediate data without costly DRAM accesses. 
For example, the \textit{kernel fusion}~\cite{Wang2010} is widely used in ML compilers and libraries. 
Similarly, algorithmic fusion, such as FlashAttention~\cite{dao2022flashattention}, reduces kernel launches and memory traffic. 
However, fusion may reduce modularity and be limited by dynamic control flow since it is workload- and compiler-dependent. 

Heterogeneous runtime environments express task dependencies and manage data movement across CPUs and GPUs. 
For example, \emph{StarPU}~\cite{Augonnet2009} uses ``codelets'' and performance models that drive dynamic scheduling and automated data coherence across memories. 
\emph{Legion}~\cite{Bauer2012} introduces a ``data-centric'' model, where tasks declare privileges over logical regions, enabling mappers and runtime to infer dependencies, place data, and generate asynchronous copies. 
Recent systems combine task runtimes with CUDA Graph to cut host dispatch overhead~\cite{Huang2022}.
The proposed SET framework is orthogonal to these approaches and can be integrated seamlessly with them.

In summary, existing techniques either optimize host-side submission, rely on coarse-grained batching, or move scheduling entirely to the device. None provides fine-grained, event-driven coordination between host and device while preserving CUDA Graph semantics. 
Most approaches operate either entirely on the host (e.g., CUDA Graph~\cite{NvidiaCudaGuide}, static batching~\cite{Ekelund2025}, queue models~\cite{Guevara2009EnablingTP}) or device (e.g., persistent kernels~\cite{Steinberger2014}).
SET fills this gap through an event-chained host–device scheduling that preserves compatibility with CUDA Graph while minimizing inter-kernel gaps and maintaining runtime flexibility.
\vspace{-3mm}
\section{Background and Motivation} \label{sec:background}
\vspace{-2mm}

\subsection{CUDA Graph}
CPU submits worker threads to GPU through a sequence of individual kernel launches and memory copy calls. Each call incurs a non-trivial amount of overhead, leading to significant bottlenecks in applications with many small operations. 
CUDA Graph is introduced to mitigate this issue by allowing a series of operations to be defined, optimized, and launched as a single unit of work~\cite{NvidiaCudaGuide}. Developers can consolidate multiple kernels into a single graph launch operation, thereby reducing the overall kernel launch overheads.

\vspace{-2mm}
\subsection{Bottleneck Analysis}
\begin{figure}[t]
    \centering
    \includegraphics[width=\linewidth]{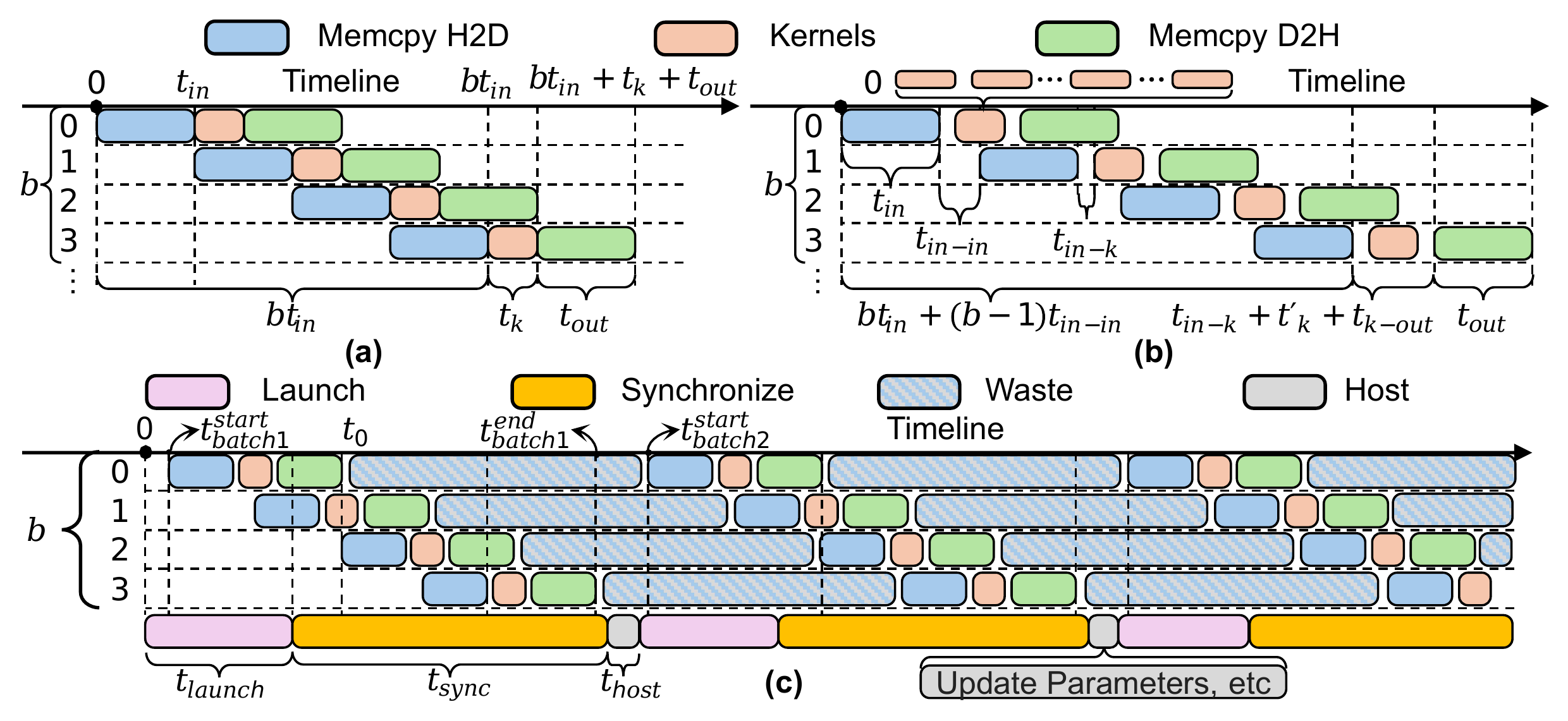}
    \vspace{-6mm}
    \caption{(a) Ideal execution flow of a static batching CUDA program (b) Execution flow with intra-batch gaps (c) Execution flow with inter-batch gaps}
    \vspace{-6mm}
    \label{fig:bottleneck}
\end{figure}

\vspace{-2mm}
We abstract general CUDA programs into three key steps: \texttt{memcpy H2D}  (host-to-device), kernels, and \texttt{memcpy D2H}  (device-to-host). 
We denote the batch size as $b$, time spent on \texttt{memcpy H2D}  as $t_{in}$, time spent on \texttt{memcpy D2H}  as $t_{out}$, time spent on kernels as $t_k$. 
As illustrated in Fig.~\ref{fig:bottleneck}(a), 
the \textit{ideal execution time} without any gaps between operations can be written as:
\begin{equation}
    T_{ideal} = b t_{in} + t_k + t_{out}
    \label{eq:ideal}
\end{equation}
During actual execution, there are two types of additional idle periods, intra-batch and inter-batch overheads as described next.

\noindent\textbf{Intra-Batch Overheads:} 
First, the host-to-device copies (\texttt{memcpy H2D}) execute with a delay between them, as denoted by $t_{in-in}$ in Fig.~\ref{fig:bottleneck}(b). 
Second, the kernels execute with a delay ($t_{in-k}$) after the input is copied. 
Third, there is delay between finishing the kernel execution and writing the output from the device to host (\texttt{memcpy D2H}), which is denoted as $t_{k-out}$ in Fig.~\ref{fig:bottleneck}(b).
Finally, suppose the total kernel execution time increases by $\Delta t_k$ due the \textit{inter-kernel} gaps. 
Then, the \textit{total intra-batch overheads} can be written as the difference between the overall execution time and $T_{ideal}$ as:
\begin{equation}
    t_{intra} = (b - 1) t_{in-in} + t_{in-k} + \Delta t_k + t_{k-out}
    \label{eq:intra}
\end{equation}

\noindent\textbf{Inter-Batch Overheads:} At the inter-batch level, the gap persists between batches, as shown in Fig.~\ref{fig:bottleneck}(c). At time $t_0$, the first job in the batch has already finished. However, the job in the next batch cannot start until $t_{batch2}^{start}$ due to synchronization and parameter updates. The preceding batch ends at $t_{batch1}^{end}$. Therefore, the \textit{inter-batch overheads} can be written as:
\begin{equation}
    t_{inter} = t_{batch2}^{start} - t_{batch1}^{end}
    \label{eq:inter}
\end{equation}

\noindent Consequently, the measured wall-clock time can be formulated in two equivalent ways: it is either the aggregate of graph launch $t_{launch}$, synchronization $t_{sync}$, and parameter update $t_{host}$; or alternatively, the sum of ideal execution time $T_{ideal}$, intra-batch overhead $t_{intra}$, and inter-batch overhead $t_{inter}$:
\begin{equation}
    \begin{aligned}
        T_{measured} &= t_{launch} + t_{sync} + t_{host} \\
        &= T_{ideal} + t_{intra} + t_{inter} \\
        &= T_{ideal} + t_{schedule}
    \end{aligned}
    \label{eq:measured}
\end{equation}

As a result, once the profiling data for kernels and operations in a program are collected, the overheads $t_{schedule}$ can be quantized with Eq.~\ref{eq:measured}. These overheads hinder developers from achieving peak GPU performance. Closing these gaps requires an event-driven, memory-safe runtime that triggers work upon per-stream completion, keeps enough jobs in flight to mask latency, and maximizes hardware utilization. SET aims to address precisely this need.

\section{Stream-Event-Triggered Scheduling}
\label{sec:methodology}

\begin{figure}[t]
    \centering
    \includegraphics[width=\linewidth]{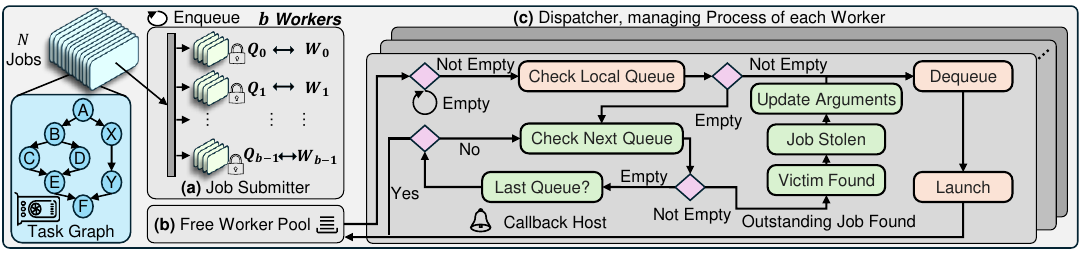}
    \vspace{-6mm}
    \caption{Overview of our runtime framework execution flow}
    \label{fig:overview}
    \vspace{-7mm}
\end{figure}

This section presents the proposed stream-event-triggered runtime framework for efficient CUDA graph pipeline scheduling. SET treats each task as a distinct CUDA graph executable. It comprises two key components: (1) the job submitter and (2) the dispatcher, as shown in Fig.~\ref{fig:overview}. The job submitter monitors job queues and inserts a new job whenever a slot becomes available. The dispatcher monitors workers' status. Once a worker becomes available, the dispatcher finds a job to feed it, either from local queue or stolen from others. A callback event is appended to the worker after each launch.

\subsection{Job-as-Graph and Memory Management}
SET abstracts tasks as DAGs, where nodes represent operations and edges denote data dependencies. The nodes are initialized with the target function, input arguments, grid and block dimensions. 
During graph construction, arguments for \texttt{memcpy} operations or kernels that vary across iterations are assigned placeholders to facilitate runtime updates.
For tasks implemented without third-party libraries (e.g., CUTLASS and cuDNN), the dependencies are defined explicitly. Otherwise, the graph is built with a combination of explicit descriptions and \texttt{streamcapture} to avoid unnecessary placeholders~\cite{Zheng2023}.

To ensure data integrity, write operations to active memory slots are prohibited. When there are multiple workers, SET keeps a batch size of $b$ memory copies because the tasks in the batch run concurrently. 
The pre-allocated memory slots are reusable across different batches since allocating new memory is costly. The stolen jobs are retargeted to the thief's buffers to guarantee memory safety without new device-side allocations.

\subsection{Event-Chained Scheduling}

\vspace{-1mm}
\subsubsection{Scheduler Components:}
SET relies on four key components: 

\noindent \textbf{(1) Workers} are the fundamental units of execution. The device is configured with $b$ workers, matching the batch size. Each worker $W_i, \forall i \in Z, 0\leq i < b$ encapsulates a dedicated CUDA stream $S_i$, a unique, pre-instantiated graph executable $G_i$ and a dedicated set of device-side buffers $M_i$. Crucially, $G_i$ is strictly bounded to operate only on the memory in $M_i$. This isolation facilitates memory-safe work-stealing. 

\noindent \textbf{(2) Per-worker job queue}. Each worker $W_i$ is assigned a thread-safe, host-side job queue $Q_i$. To mitigate the overhead of on-the-fly argument updates, $Q_i$ stores fully prepared graph executables rather than simple task indices. 

\noindent \textbf{(3) Free worker pool}. The thread-safe queue $W_{\mathrm{pool}}$ maintains the indices of currently available workers. Rather than employing polling or round-robin checks, the pool is updated only upon task completion, thereby reducing host-device communication. 

\noindent \textbf{(4) Submitter and Dispatcher}. These 
are two distinct host threads. The submitter serves as a \textit{producer}. It continuously monitors the job queues of all workers, prepares the graph executable by updating its parameters and determines when to add the new job and which queue to add it to. The dispatcher serves as a \textit{consumer}. It monitors whether there is an available worker. If there is, it fetches a job to feed it. They work closely together to ensure that all workers are fed quickly without oversubscribing host-device synchronization. 

\vspace{-2mm}
\subsubsection{Initialization:} 
The main thread 
(1) populates all per-worker job queues \\
$\{Q_0, \ldots, Q_b\}$, and 
(2) initializes all workers $\{W_0, \ldots, W_b\}$  with their corresponding streams, buffers, and graph executables. 
Then, it spawns the job submitter and dispatcher threads. 

\vspace{-2mm}
\subsubsection{Job Submission} 
The job submitter monitors all queues for available slots. Upon finding a free slot, it proceeds to acquire the corresponding queue, updates the parameters for job $J$ and enqueues it, as detailed in Algorithm~\ref{alg:submitter} (lines 3--5) and Fig.~\ref{fig:overview}(a). 
The jobs within a selected queue $Q_i, \forall i \in Z, 0\leq i < b$ are expected to be executed by worker $W_i$. 
This one-to-one correspondence allows their parameters to be updated for the worker before insertion.

\begin{figure}[t]
\centering
\begin{minipage}[t]{0.38\linewidth}
\begin{algorithm}[H]
\setcounter{algocf}{0}
\caption{Job Submitter}
\label{alg:submitter}
\SetAlgoNoEnd
\small
\KwData{Next job counter $c_{\mathrm{next}}$, job queue $Q_i$, job $J$}
\KwResult{Populated job queue $Q_i~\forall i \in Z, 0\leq i < b$}

id $\leftarrow$ $c_{\mathrm{next}}$; \\
\If{$Q_{i}$ not full}{
$c_{\mathrm{next}}$ $\leftarrow$ compare\_exchange($c_{\mathrm{next}}$, id, id+1); \\
UpdateGraphParams($J_\mathrm{id}$, $i$); \\
$Q_i$.push($J_\mathrm{id}$); \\
}
\end{algorithm}

\vspace{-0.2em}

\begin{algorithm}[H]
\setcounter{algocf}{2}
\caption{Asynchronous Resource Return (Callback)}
\label{alg:callback}
\small
\KwData{Done job counter $c_\mathrm{done}$, completed worker $w_{id}$}
\KwResult{Freed worker returned to pools}

$c_{\mathrm{done}}$.atomic\_fetch\_add(1)\; 
$W_\mathrm{pool}$.push($w_{id}$)\;
notify\_one(); \\
\end{algorithm}
\end{minipage}\hfill
\begin{minipage}[t]{0.60\linewidth}
\begin{algorithm}[H]
\setcounter{algocf}{1}
\caption{Job Dispatch \& Launch}
\label{alg:merged_dispatcher}
\SetAlgoNoEnd
\small
% --- Input/Output ---
\KwData{Free worker pool $W_\mathrm{pool}$, job queues $Q_i$}
\KwResult{Dispatched GPU jobs}

% --- Function Setup ---
\SetKwFunction{FindJob}{FindJob}
\SetKwFunction{LaunchJob}{LaunchJob}
\SetKwProg{Fn}{Function}{}{}

% --- Main Dispatcher Logic ---
\tcp{Dispatcher:}
\If{$W_\mathrm{pool}$ not empty}{
    $w_{id} \leftarrow W_\mathrm{pool}$.pop()\;
    
    $job \leftarrow \FindJob(w_{id}, Q_i)$\;
    
    \If{$job \neq \mathrm{NULL}$}{
        \LaunchJob($job$, $w_{id}$)\;
    } \Else {
        $Q_W$.push($w_{id}$)\;
    }
}

\vspace{0.5em}

% --- Function 1: FindJob ---
\Fn{\FindJob{$w_{id}$, $Q_i$}}{%
    \If{$Q_{w_{id}}.\text{try\_pop}(job) = \text{true}$}{%
        $job.is\_stolen \leftarrow \text{false}$\;
        \KwRet $job$\;
    }
    
    \For{$k \leftarrow 1$ \KwTo $b-1$}{%
        $victim\_id \leftarrow (w_{id} + k) \bmod b$\;
        \If{$Q_{victim\_id}.\text{try\_steal}(job) = \text{true}$}{%
            $job.is\_stolen \leftarrow \text{true}$\;
            \KwRet $job$\;
        }
    }
    \KwRet $\text{NULL}$\;
}
\vspace{0.5em}

% --- Function 2: LaunchJob ---
\Fn{\LaunchJob{$job$, $w_{id}$}}{
    \If{$job.is\_stolen = \text{true}$}{%
        UpdateGraphParams(job, $w_{id}$)\;%
    }
    GraphLaunch($job$, $w_{id}$); \\
    AddCallback($job$, $w_{id}$, Callback); \\
    \KwRet\;
}
\end{algorithm}
\end{minipage}
\vspace{-5mm}
\end{figure}

\vspace{-2mm}
\subsubsection{Job Dispatch and Launch}
In parallel with \textit{job submission}, the dispatcher thread monitors the free worker pool $W_{\mathrm{pool}}$, as shown in Fig.~\ref{fig:overview}(b). 
Upon retrieving an idle worker $W_\mathrm{free}$, the dispatcher invokes the \texttt{FindJob} function to search for a ready job, as in Algorithm~\ref{alg:merged_dispatcher} (lines 2--3).  
While searching, the dispatcher first attempts to acquire a job from the selected worker's ($W_\mathrm{free}$) dedicated queue ($Q_{\mathrm{free}}$). 
If the local queue is not empty, the head of $Q_{\mathrm{free}}$ is popped and launched.
Otherwise, the dispatcher invokes a work-stealing policy by iterating through peer queues $Q_k$, where $k \neq \mathrm{free}$ and attempting to steal the first job it meets, as shown in Algorithm~\ref{alg:merged_dispatcher} (lines 9--17) and Fig.~\ref{fig:overview}(c).  

The next step is launching the job (\texttt{LaunchJob} on line 5 in Algorithm~\ref{alg:merged_dispatcher}). 
If the job was found in the free worker's local queue, then the dispatcher directly launches the job on worker $W_\mathrm{free}$.
Otherwise (i.e., it was stolen from another queue), 
the graph executable is natively bound to the victim's memory resources. To address this mismatch, a just-in-time (JIT) update is performed to retarget the graph parameters. This update rebinds the executable to (1) read from $W_\mathrm{free}$'s input memory, (2) use $W_\mathrm{free}$'s buffers and (3) write to $W_\mathrm{free}$'s output memory. Following the reconfiguration, the dispatcher launches the ``stolen'' job on $w_{\mathrm{free}}$, as shown in Fig.~\ref{fig:overview}(c) and Algorithm~\ref{alg:merged_dispatcher} (lines 19--23).

% \begin{algorithm}[ht]
% \caption{Job Dispatcher and Launch}
% \label{alg:merged_dispatcher}
% \SetAlgoNoEnd

% % --- Input/Output ---
% \KwData{Free worker pool $W_\mathrm{pool}$, job queues $Q_i$}
% \KwResult{Dispatched GPU jobs}

% % --- Function Setup ---
% \SetKwFunction{FindJob}{FindJob}
% \SetKwFunction{LaunchJob}{LaunchJob}
% \SetKwProg{Fn}{Function}{}{}

% % --- Main Dispatcher Logic ---
% \tcp{Dispatcher:}
% \If{$W_\mathrm{pool}$ not empty}{
%     $w_{id} \leftarrow W_\mathrm{pool}$.pop()\;
    
%     $job \leftarrow \FindJob(w_{id}, Q_i)$\;
    
%     \If{$job \neq \mathrm{NULL}$}{
%         \LaunchJob($job$, $w_{id}$)\;
%     } \Else {
%         $Q_W$.push($w_{id}$)\;
%     }
% }

% \vspace{0.5em}

% % --- Function 1: FindJob ---
% \Fn{\FindJob{$w_{id}$, $Q_i$}}{%
%     \If{$Q_{w_{id}}.\text{try\_pop}(job) = \text{true}$}{%
%         $job.is\_stolen \leftarrow \text{false}$\;
%         \KwRet $job$\;
%     }
    
%     \For{$k \leftarrow 1$ \KwTo $b-1$}{%
%         $victim\_id \leftarrow (w_{id} + k) \bmod b$\;
%         \If{$Q_{victim\_id}.\text{try\_steal}(job) = \text{true}$}{%
%             $job.is\_stolen \leftarrow \text{true}$\;
%             \KwRet $job$\;
%         }
%     }
%     \KwRet $\text{NULL}$\;
% }
% \vspace{0.5em}

% % --- Function 2: LaunchJob ---
% \Fn{\LaunchJob{$job$, $w_{id}$}}{
%     \If{$job.is\_stolen = \text{true}$}{%
%         UpdateGraphParams(job, $w_{id}$)\;%
%     }
%     GraphLaunch($job$, $w_{id}$); \\
%     AddCallback($job$, $w_{id}$, Callback); \\
%     \KwRet\;
% }
% \end{algorithm}

\vspace{-2mm}
\subsubsection{Asynchronous Resource Return}
After each launch, whether that job was stolen or not, the dispatcher appends an asynchronous callback event to worker $W_\mathrm{free}$ at the end of the stream. When $W_\mathrm{free}$ completes all operations in the graph, the callback event is triggered by the CUDA driver on a separate thread. This event atomically increments the counter for finished jobs and returns $W_\mathrm{free}$ to the $W_{\mathrm{pool}}$. This mechanism automatically recycles the worker, making it available for subsequent iterations of the dispatch loop, as detailed in Algorithm~\ref{alg:callback}.

\section{Experimental Evaluations} \label{sec:evaluation}
\vspace{-1mm}

\subsection{Evaluation Setup} \label{sec:setup}

\subsubsection{Hardware Configuration}
All evaluations are conducted on two Linux servers, one equipped with Intel Xeon Gold 6330, NVIDIA RTX 3090 and 192 GB RAM, the other equipped with Intel I7-11700, NVIDIA RTX 5090 and 128 GB RAM. The servers run on Ubuntu 22.04, with NVIDIA CUDA v12.8. All source code is compiled with \texttt{O2} optimization and C++20 standard.
\vspace{-4mm}

\subsubsection{Benchmark Applications}

\begin{figure}[t]
    \centering
    \includegraphics[width=0.7\linewidth]{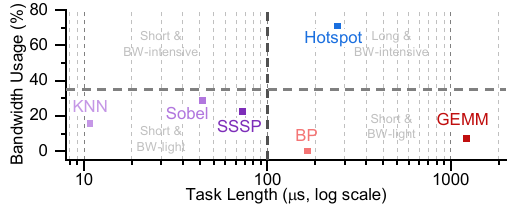}
    \vspace{-4mm}
    \caption{Average memory usage and task lengths of benchmarks used in this work} \label{fig:workloads_scatter}
    \vspace{-6mm}
\end{figure}

We employ six workloads that span compute- and memory-bound behavior, as shown in Fig.~\ref{fig:workloads_scatter}.

\begin{enumerate}[leftmargin=*]
\item \textbf{Sobel operator}~\cite{Sobel2015} applies normalization, edge detection, mean filtering, binary thresholding, and blending operation to input images.

\item \textbf{General matrix multiplication (GEMM)} performs tile-based dense matrix multiplication.

\item \textbf{Back propagation (BP)}~\cite{Mitchell2013} workload performs a single-layer training step with on-device synthetic minibatch generation. 

\item \textbf{K-nearest neighbor (KNN)}~\cite{Guo2003}) workload performs supervised classification via brute-force search for feature vectors. 

\item \textbf{Hotspot}~\cite{Huang2006} runs iterative thermal simulations solving differential equations.

\item \textbf{Single-source shortest path (SSSP)}~\cite{Bellman1958} workload performs Bellman–Ford graph traversal with frontier-based relaxation for improved GPU parallelism.
\end{enumerate}

We employ throughput as the primary performance metric and analyze the scheduling overheads using NVIDIA Nsight Systems Profiler.

\vspace{-1mm}
\subsubsection{Baseline Models}
We compare the proposed SET framework against four representative programming models, which are 
implemented with best-practice optimizations to ensure fairness.

\begin{enumerate}[leftmargin=*]
    \item \textbf{Synchronous model}~\cite{NvidiaCudaGuide} launches kernels one by one on the same stream without any explicit scheduling mechanism.

    \item \textbf{Graph model}~\cite{NvidiaCudaGuide} pre-instantiates the job as a graph, then launches it repeatedly.
    We reduced data movement overheads when copying input values to graph's placeholders~\cite{Zheng2023} using explicit API calls and stream capturing. 
    
    \item \textbf{Static batching model}~\cite{Ekelund2025}. While maintaining the batching model's scheduler, we adopt dynamic CUDA graph construction to enable on-the-fly parameter updates. 
    
    \item \textbf{Queue model}~\cite{Guevara2009EnablingTP}. We keep the queue model's scheduler ``issue queue'' unchanged, and change its job wrapper from merged kernels to CUDA graphs to make sure it holds the same input as other baseline models.

\end{enumerate}

\subsection{Performance (Throughput) Evaluation}

\begin{figure*}[ht]
    \centering
    \includegraphics[width=\linewidth]{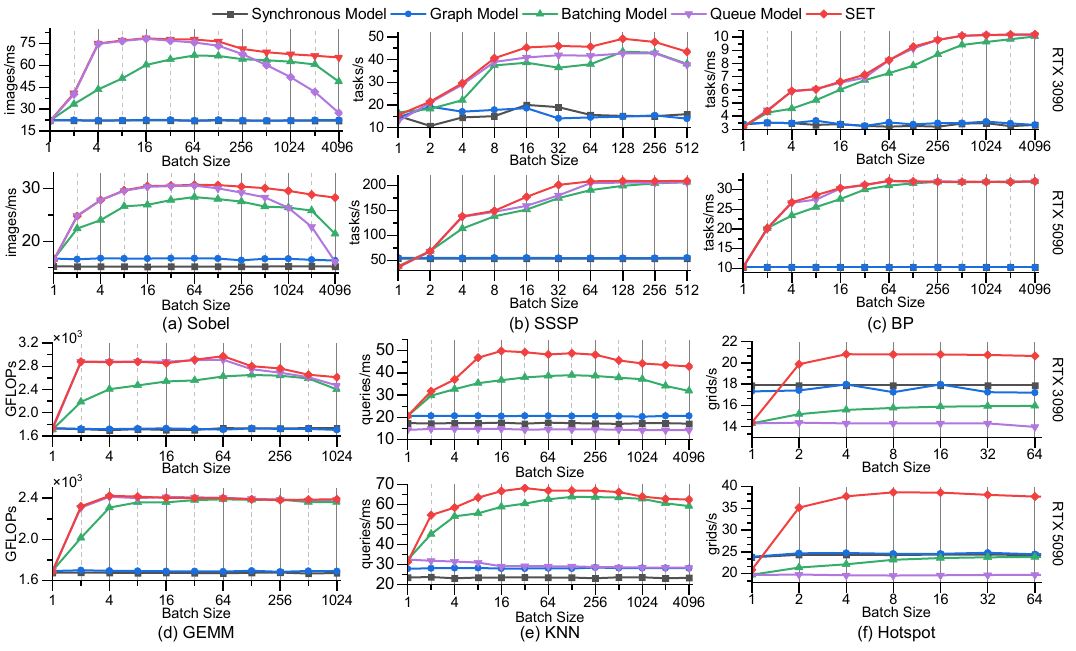}
    \vspace{-8mm}
    \caption{Throughput vs. batch size in (a) Sobel (img/ms), (b) GEMM (GFLOPs), (c) BP (tasks/s), (d) KNN (queries/ms), (e) Hotspot (grids/s), and (f) SSSP (tasks/s) workloads.}
    \label{fig:six_perf_vs_bs}
    \vspace{-6mm}
\end{figure*}

Fig.~\ref{fig:six_perf_vs_bs} plots the throughput as a function of the batch size $b$ across all workloads and programming models. 
GEMM, Hotspot, and SSSP have a lower maximum batch size since they exceed the memory capacity with larger batch sizes.

\noindent\textbf{General Trends:} 
Throughput increases with the batch size since the GPU can run more tasks in parallel and maximize hardware utilization. 
For all workloads, SET, queue model, and batching models achieve higher throughput as the batch size grows until reaching maximum hardware utilization. As the batch size continues to grow, the scheduling overheads become dominant, as explained in Section~\ref{sec:scheduling_overheads}.
In contrast, the synchronous and graph models do not benefit from batching and appear as flat lines since they only have one worker stream. Therefore, their performance is limited by the host latency to emit jobs. 
They cannot utilize all hardware resources. 
Consequently, they have significantly lower throughput than other approaches except for Hotspot workload, as detailed under the workload-specific analysis.

\noindent\textbf{Sobel, SSSP, BP, GEMM Analysis (Fig.~\ref{fig:six_perf_vs_bs}(a)-(d)):}
SET outperforms the synchronous and graph models with a significant margin, achieving 2.31$\times$ and 2.27$\times$ speedup on average, respectively, as summarized in Table~\ref{tab:comparison}. 
Unlike the synchronous and graph models, the batching model benefits from larger batch sizes and achieves a higher throughput. 
However, the inter-batch level overheads become the bottleneck, thus degrading its throughput.
As a result, SET achieves 1.10$\times$ speedup over the batching model on average.
The queue model has the closest performance to the proposed SET framework under these workloads. 
It achieves the same performance for batch sizes up to 32 in Sobel, 4 in SSSP, 4096 in BP, and 32 in GEMM. 
Beyond the thresholds, SET outperforms the queue model due to lower scheduling overheads when kernels saturate the hardware, as detailed in Section~\ref{sec:scheduling_overheads}. The queue model oversubscribes the bandwidth, exacerbating its performance degradation in Sobel, where kernels involve over 30\% L2 cache traffic.
On average, SET still achieves 1.06$\times$ speedup over the queue model.

\noindent\textbf{The KNN Workload Fig.~\ref{fig:six_perf_vs_bs}(e):} 
We analyze the KNN workload separately since it exhibits a different behavior, 
where the queue model performs poorly even worse than the synchronous model. As shown in Fig.~\ref{fig:workloads_scatter}, the KNN workload involves many kernels whose execution time is as small as an average of \SI{10}{\micro \second}, the same order to acquire and release thread lock (usually around \SI{3}{\micro \second} to \SI{5}{\micro \second} when facing data contention). 
In the queue model, many small kernels invoke the host to acquire/release a mutex to its global queue frequently, 
while workers are waiting for the scheduler to dispatch them a job. Therefore, its performance is even worse than the synchronous and graph models and cannot improve with larger batch sizes. 
In contrast, the proposed SET framework still holds a high throughput in KNN workload thanks to its per-worker job queue and ready-to-go job wrapper, eliminating the frequent mutex acquire/release and the as-needed updates to graph nodes' arguments, which become important when dealing with small kernels. 

\noindent\textbf{The Hotspot Workload (Fig.~\ref{fig:six_perf_vs_bs}(f)):} 
Another extreme workload is Hotspot, which is a heavily memory-bound workload as shown in Fig.~\ref{fig:workloads_scatter}. 
Profiling data from Nsight Compute shows that Hotspot uses up to 90\% of DRAM bandwidth and up to 50\% of L2 cache traffic. 
So, the GPU throughput can easily saturate when the DRAM reaches its maximum utilization, as more kernels just split the same fixed bandwidth across more requests. The queue and batching models suffer from high data contention since they launch more jobs than the device can hold efficiently. In this case, the synchronous model and graph model even perform better than them, since they only process one job at a time, which puts less pressure on TLB, L2 cache and DRAM and has higher spatial locality and temporal locality. 
In contrast, SET still benefits from launching the jobs on demand and overlapping device copies with kernels. 
As shown in Fig.~\ref{fig:six_perf_vs_bs}(f), SET is the only technique that benefits from larger batch sizes and outperforms all baselines by a significant margin.

\noindent\textbf{Throughput Comparison Summary:}
SET delivers the best throughput compared to other models, achieving an average (average on RTX 3090 and 5090) of 2.18$\times$, 2.1$\times$, 1.17$\times$, 1.39$\times$ speedup against synchronous, graph, batching and queue models, respectively (see Table~\ref{tab:comparison}). These results, aggregated across RTX 3090 and 5090 GPUs, demonstrate that SET maintains high performance even under demanding workloads characterized by small kernels and high memory pressure (Fig.~\ref{fig:workloads_scatter}, Fig.~\ref{fig:six_perf_vs_bs}). Such improvements underscore the robustness of our approach across diverse execution environments.

\begin{table}[t]
    \centering
    \caption{Speedups over baseline models on RTX 3090 (Ampere architecture) and RTX 5090 (Blackwell architecture)}
    \label{tab:comparison}
    \begin{tabular}{lcccccccc}
    \toprule
    \multirow{2}{*}{Speedup} & \multicolumn{2}{c}{Synchronous} & \multicolumn{2}{c}{Graph} & \multicolumn{2}{c}{Batching} & \multicolumn{2}{c}{Queue} \\
    \cmidrule(lr){2-3} \cmidrule(lr){4-5} \cmidrule(lr){6-7} \cmidrule(lr){8-9}
    & Ampere & Blackwell & Ampere & Blackwell & Ampere & Blackwell & Ampere & Blackwell \\
    \midrule
    Sobel  & 2.99$\times$ & 1.86$\times$ & 2.97$\times$ & 1.71$\times$ & 1.23$\times$ & 1.12$\times$ & 1.20$\times$ & 1.10$\times$ \\
    SSSP  & 2.45$\times$ & 3.07$\times$ & 2.46$\times$ & 2.99$\times$ & 1.15$\times$ & 1.07$\times$ & 1.10$\times$ & 1.03$\times$ \\
    BP    & 2.34$\times$ & 2.78$\times$ & 2.26$\times$ & 2.78$\times$ & 1.10$\times$ & 1.04$\times$ & 1.01$\times$ & 1.01$\times$ \\
    GEMM  & 1.58$\times$ & 1.39$\times$ & 1.58$\times$ & 1.38$\times$ & 1.12$\times$ & 1.02$\times$ & 1.01$\times$ & 1.01$\times$ \\
    KNN   & 2.47$\times$ & 2.63$\times$ & 2.08$\times$ & 2.19$\times$ & 1.23$\times$ & 1.08$\times$ & 2.94$\times$ & 2.09$\times$ \\
    Hotspot & 1.10$\times$ & 1.47$\times$ & 1.39$\times$ & 1.45$\times$ & 1.27$\times$ & 1.56$\times$ & 1.38$\times$ & 1.81$\times$ \\
    \midrule
    Average & 2.15$\times$ & 2.20$\times$ & 2.12$\times$ & 2.08$\times$ & 1.18$\times$ & 1.15$\times$ & 1.44$\times$ & 1.34$\times$ \\
    \bottomrule
    \end{tabular}
    \vspace{-3mm}
\end{table}

\subsection{Scheduling Overhead Analysis} \label{sec:scheduling_overheads}

\begin{figure}[htbp]
     \centering
     \includegraphics[width=\linewidth]{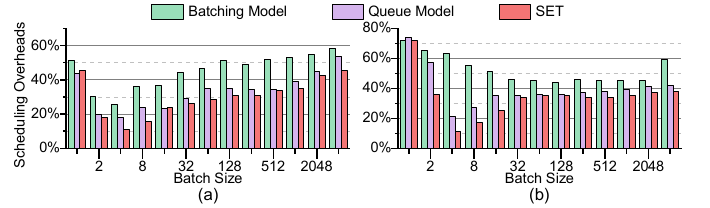}
     \vspace{-8mm}
     \caption{Scheduling overheads with different batch sizes in different models. (a) on RTX 3090 (b) on RTX 5090}
     \vspace{-8mm}
     \label{fig:overheads_share}
\end{figure}

Fig.~\ref{fig:overheads_share} plots the scheduling overheads over the total execution time for the three best performing programming models: batching, queue model and SET. As shown in Eq.~(\ref{eq:measured}), the fraction is calculated by $\mathrm{Fraction} = \frac{t_{schedule}}{T_{measured}}$. 

All workloads have low hardware utilization when the batch size equals one since the launch overheads are high relative to the kernel execution time.
Specifically, scheduling overhead accounts for 45-51\% of the total execution time on RTX 3090, resulting in low throughput. This behavior is an artifact of inter-kernel gaps and consistent with the analysis in Section~\ref{sec:background} and the execution flow shown in Fig.~\ref{fig:bottleneck}(b). 
Throughput increases with the batch size, as more kernels can collectively saturate the GPU, thereby reducing the relative contribution of scheduling overhead. Indeed, the scheduling overhead of the SET, queue, and batch models drops to 11\%, 18\%, and 25\% on RTX 3090, respectively, when the batch size is four. 

As the batch size grows, the inter-batch overheads become significant. This effect is most pronounced in the batching model, as it synchronizes across batches at the end of each batch. As a result, its scheduling overheads increase from 36\% with a batch size of 8 to 59\% when the batch size is 4096. 
The queue model and our method incur host-side overheads (e.g., job queues and free worker pool monitoring). These costs remain lower than the batching model and are negligible when workers do not cause high data contention. However, the queue model suffers more from managing shared resources than SET, which incurs substantial contention on its global mutex, resulting in up to 54\% overhead at a batch size of 4096. In contrast, SET alleviates this bottleneck by using per-worker queues and minimizing shared-queue operations, resulting in lower overhead at high batch sizes. Overall, SET achieves consistently lower scheduling overheads (15\% with a batch size of 8 and 45\% with a batch size of 4096). 
On average, SET has 54.64\% and 18.62\% lower scheduling overhead than the batch and queue models.

\vspace{-6mm}
\begin{table}[ht]
\centering
\caption{Comparison of Scheduling Overheads on RTX 3090 and RTX 5090}
\label{tab:avg_overhead_comparison}
\begin{tabular}{lccc}
\toprule
Average Ratio & Batching & Queue & SET \\ 
\midrule
RTX 3090 (Ampere) & 45.32\%  & 33.36\% & 29.83\% \\ 
RTX 5090 (Blackwell) & 52.38\% & 39.85\% & 34.62\% \\
\bottomrule
\end{tabular}
\vspace{-6mm}
\end{table}

While the previous analysis focused on the Ampere architecture (RTX 3090), these overheads persist, and are often exacerbated on Blackwell architecture (RTX 5090). Our profiling indicates that moving to Blackwell yields a $1.79\times$ average speedup in kernel execution time ($T_{\mathrm{compute}}$), driven by increased CUDA core counts, higher clock frequencies, and improved IPC (instruction-per-cycle). However, host-side scheduling ($T_{\mathrm{overhead}}$) for operations such as graph instantiation and parameter updates scales more slowly than raw compute throughput. In accordance with Amdahl's Law, as $T_{\mathrm{compute}}$ diminishes, $T_{\mathrm{overhead}}$ exerts a greater influence on total execution time. As shown in Table~\ref{tab:avg_overhead_comparison}, the relative overhead ratio is higher on the RTX 5090, confirming that modern GPU architectures are increasingly bottlenecked by scheduling inefficiencies.

In summary, the scheduling overheads are large with small batches due to low hardware utilization. They reduce with higher GPU utilization and reach a minimum value. However, they increase again due to inter-batch overheads, resulting in a U-shaped behavior. This behavior explains the inverted U-shaped throughput curves observed in Sobel, GEMM, and SSSP in Fig.~\ref{fig:six_perf_vs_bs}. 

\section{Conclusions} \label{sec:conclusion}
\vspace{-2mm}

This paper presents SET, a stream-event-triggered scheduling framework for efficient CUDA graph pipelines. The analytical model clearly shows that the scheduling overheads consist of two parts, intra-batch overheads and inter-batch overheads. To minimize the overheads, SET maintains a job submitter to feed the job queues and a dispatcher to fetch a job for any available worker. SET implements an event-driven callback signal to enable asynchronous resource release upon a job completion. We evaluate its performance on six real-world workloads, where SET achieves 1.15--1.44$\times$ speedup over the state-of-the-art baselines and reduces the scheduling overheads by 18-54\%. SET provides an orthogonal, kernel-agnostic design and is friendly to other optimizations to kernels or compilers.
\vspace{-8mm}

\subsubsection{Disclosure of Interests} Dr. Ogras serves as a contractor for Samsung Austin Research \& Development Center and Advanced Computing Lab (SARC/ACL). This relationship has been approved under applicable outside activities policies.

\bibliographystyle{splncs04}
\bibliography{reference.bib}

@InProceedings{Ekelund2025,
  author    = {Ekelund, Jonah and Markidis, Stefano and Peng, Ivy},
  booktitle = {2025 33rd Euromicro International Conference on Parallel, Distributed, and Network-Based Processing (PDP)},
  title     = {Boosting Performance of Iterative Applications on GPUs: Kernel Batching with CUDA Graphs},
  year      = {2025},
  month     = mar,
  pages     = {70--77},
  publisher = {IEEE},
  doi       = {10.1109/pdp66500.2025.00019},
}

@InProceedings{Zheng2023,
  author     = {Zheng, Bojian and others},
  booktitle  = {56th Annual IEEE/ACM International Symposium on Microarchitecture},
  title      = {Grape: Practical and Efficient Graphed Execution for Dynamic Deep Neural Networks on GPUs},
  year       = {2023},
  month      = oct,
  pages      = {1364--1380},
  publisher  = {ACM},
  series     = {MICRO ’23},
  collection = {MICRO ’23},
  doi        = {10.1145/3613424.3614248},
}

@Misc{Sobel2015,
  author    = {Sobel, Irwin and Feldman, Gary},
  title     = {An Isotropic 3x3 Image Gradient Operator},
  year      = {2015},
  doi       = {10.13140/RG.2.1.1912.4965},
  language  = {en},
  publisher = {Unpublished},
}

@Misc{NvidiaCudaGuide,
  author       = {NVIDIA Corp.},
  howpublished = {\url{https://docs.nvidia.com/cuda/cuda-c-programming-guide/index.html}},
  month        = {Oct},
  note         = {Accessed: Nov. 3, 2025},
  title        = {CUDA C++ Programming Guide},
  year         = {2025},
}

@InProceedings{Guevara2009EnablingTP,
  author = {Marisabel Guevara and others},
  title  = {Enabling Task Parallelism in the CUDA Scheduler},
  year   = {2009},
  url    = {https://api.semanticscholar.org/CorpusID:306206},
}

@InBook{Guo2003,
  author    = {Guo, Gongde and Wang, Hui and Bell, David and Bi, Yaxin and Greer, Kieran},
  pages     = {986--996},
  publisher = {Springer Berlin Heidelberg},
  title     = {KNN Model-Based Approach in Classification},
  year      = {2003},
  isbn      = {9783540399643},
  booktitle = {On The Move to Meaningful Internet Systems 2003: CoopIS, DOA, and ODBASE},
  doi       = {10.1007/978-3-540-39964-3_62},
  issn      = {1611-3349},
}

@Article{Huang2006,
  author    = {Wei Huang and others},
  journal   = {IEEE Transactions on Very Large Scale Integration (VLSI) Systems},
  title     = {HotSpot: a compact thermal modeling methodology for early-stage VLSI design},
  year      = {2006},
  issn      = {1557-9999},
  month     = may,
  number    = {5},
  pages     = {501--513},
  volume    = {14},
  doi       = {10.1109/tvlsi.2006.876103},
  publisher = {Institute of Electrical and Electronics Engineers (IEEE)},
}

@Article{Bellman1958,
  author    = {Bellman, Richard},
  journal   = {Quarterly of Applied Mathematics},
  title     = {On a routing problem},
  year      = {1958},
  issn      = {1552-4485},
  month     = apr,
  number    = {1},
  pages     = {87--90},
  volume    = {16},
  doi       = {10.1090/qam/102435},
  publisher = {American Mathematical Society (AMS)},
}

@Book{Mitchell2013,
  author    = {Mitchell, Tom M.},
  publisher = {McGraw-Hill},
  title     = {Machine learning},
  year      = {2013},
  address   = {New York [u.a.]},
  edition   = {[Nachdr.]},
  isbn      = {0071154671},
  series    = {McGraw-Hill international editions},
  pagetotal = {414},
  ppn_gvk   = {1609271009},
}

@InProceedings{Wang2010,
  author    = {Wang, Guibin and Lin, YiSong and Yi, Wei},
  booktitle = {2010 IEEE/ACM International Conference on Green Computing and Communications \&amp; International Conference on Cyber, Physical and Social Computing},
  title     = {Kernel Fusion: An Effective Method for Better Power Efficiency on Multithreaded GPU},
  year      = {2010},
  month     = dec,
  pages     = {344--350},
  publisher = {IEEE},
  doi       = {10.1109/greencom-cpscom.2010.102},
}

@inproceedings{
dao2022flashattention,
title={FlashAttention: Fast and Memory-Efficient Exact Attention with {IO}-Awareness},
author={Tri Dao and others},
booktitle={Advances in Neural Information Processing Systems},
editor={Alice H. Oh and Alekh Agarwal and Danielle Belgrave and Kyunghyun Cho},
year={2022},
url={https://openreview.net/forum?id=H4DqfPSibmx}
}

@InBook{Augonnet2009,
  author    = {Augonnet, Cedric and others},
  pages     = {863--874},
  publisher = {Springer Berlin Heidelberg},
  title     = {StarPU: A Unified Platform for Task Scheduling on Heterogeneous Multicore Architectures},
  year      = {2009},
  isbn      = {9783642038693},
  booktitle = {Euro-Par 2009 Parallel Processing},
  doi       = {10.1007/978-3-642-03869-3_80},
  issn      = {1611-3349},
}

@InProceedings{Bauer2012,
  author    = {Bauer, Michael and Treichler, Sean and Slaughter, Elliott and Aiken, Alex},
  booktitle = {2012 International Conference for High Performance Computing, Networking, Storage and Analysis},
  title     = {Legion: Expressing locality and independence with logical regions},
  year      = {2012},
  month     = nov,
  publisher = {IEEE},
  doi       = {10.1109/sc.2012.71},
}

@Article{Huang2022,
  author    = {Huang, Tsung-Wei and Lin, Dian-Lun and Lin, Chun-Xun and Lin, Yibo},
  journal   = {IEEE Transactions on Parallel and Distributed Systems},
  title     = {Taskflow: A Lightweight Parallel and Heterogeneous Task Graph Computing System},
  year      = {2022},
  issn      = {2161-9883},
  month     = jun,
  number    = {6},
  pages     = {1303--1320},
  volume    = {33},
  doi       = {10.1109/tpds.2021.3104255},
  publisher = {Institute of Electrical and Electronics Engineers (IEEE)},
}

@Article{Steinberger2014,
  author    = {Steinberger, Markus and others},
  journal   = {ACM Transactions on Graphics},
  title     = {Whippletree: task-based scheduling of dynamic workloads on the GPU},
  year      = {2014},
  issn      = {1557-7368},
  month     = nov,
  number    = {6},
  pages     = {1--11},
  volume    = {33},
  doi       = {10.1145/2661229.2661250},
  publisher = {Association for Computing Machinery (ACM)},
}

@Article{Zhong2014,
  author    = {Zhong, Jianlong and He, Bingsheng},
  journal   = {IEEE Transactions on Parallel and Distributed Systems},
  title     = {Kernelet: High-Throughput GPU Kernel Executions with Dynamic Slicing and Scheduling},
  year      = {2014},
  issn      = {1045-9219},
  month     = jun,
  number    = {6},
  pages     = {1522--1532},
  volume    = {25},
  doi       = {10.1109/tpds.2013.257},
  publisher = {Institute of Electrical and Electronics Engineers (IEEE)},
}

@Article{Steinberger2012,
  author    = {Steinberger, Markus and Kainz, Bernhard and Kerbl, Bernhard and Hauswiesner, Stefan and Kenzel, Michael and Schmalstieg, Dieter},
  journal   = {ACM Transactions on Graphics},
  title     = {Softshell: dynamic scheduling on GPUs},
  year      = {2012},
  issn      = {1557-7368},
  month     = nov,
  number    = {6},
  pages     = {1--11},
  volume    = {31},
  doi       = {10.1145/2366145.2366180},
  publisher = {Association for Computing Machinery (ACM)},
}
\end{document}